\documentclass[12pt,preprint]{aastex}

\shorttitle{Improved Physical Parameters of TrES-2}
\shortauthors{Sozzetti et al.}

\begin{document}


\title{Improving Stellar and Planetary Parameters of Transiting Planet 
Systems: The Case of TrES-2}


\author{Alessandro Sozzetti\altaffilmark{1,2}, 
Guillermo Torres\altaffilmark{1}, David
Charbonneau\altaffilmark{1,6}, David W.\ Latham\altaffilmark{1}, 
Matthew J.\ Holman\altaffilmark{1}, 
Joshua N.\ Winn\altaffilmark{3}, John B.\ Laird\altaffilmark{4}, 
and Francis T.\ O'Donovan\altaffilmark{5}} 
\altaffiltext{1}{Harvard-Smithsonian Center for Astrophysics, 60
Garden Street, Cambridge, MA 02138 USA; asozzett@cfa.harvard.edu}
\altaffiltext{2}{INAF - Osservatorio Astronomico di Torino, 10025 Pino
Torinese, Italy}
\altaffiltext{3}{Department of Physics, and Kavli Institute for Astrophysics 
and Space Research, Massachusetts Institute of Technology, Cambridge, MA 02139 USA}
\altaffiltext{4}{Department of Physics \& Astronomy,
Bowling Green State University, Bowling Green, OH 43403 USA}
\altaffiltext{5}{California Institute of Technology, 
1200 East California Boulevard, Pasadena, CA 91125}
\altaffiltext{6}{Alfred P.\ Sloan Research Fellow}





\begin{abstract}

We report on a spectroscopic determination of the atmospheric
parameters and chemical abundance of the parent star of the recently
discovered transiting planet \mbox{TrES-2}.  A detailed LTE analysis
of a set of \ion{Fe}{1} and \ion{Fe}{2} lines from our Keck spectra
yields $T_\mathrm{eff} = 5850\pm 50$ K, $\log g = 4.4\pm 0.1$, and
[Fe/H] $= -0.15\pm 0.10$. Several independent checks (e.g., additional
spectroscopy, line-depth ratios) confirm the reliability of our
spectroscopic $T_\mathrm{eff}$ estimate.  The mass and radius of the
star, needed to determine the properties of the planet, are
traditionally inferred by comparison with stellar evolution models
using $T_\mathrm{eff}$ and some measure of the stellar luminosity,
such as the spectroscopic surface gravity (when a trigonometric
parallax is unavailable, as in this case). We apply here a new method
in which we use instead of $\log g$ the normalized separation
$a/R_\star$ (related to the stellar density), which can be determined
directly from the light curves of transiting planets with much greater
precision. With the $a/R_\star$ value from the light curve analysis of
Holman et al.~\citeyearpar{holman07b} and our $T_\mathrm{eff}$
estimate we obtain $M_\star = 0.980\pm0.062~M_\odot$ and $R_\star =
1.000_{-0.033}^{+0.036}~R_\odot$, and an evolutionary age of
$5.1^{+2.7}_{-2.3}$ Gyr, in good agreement with other constraints
based on the strength of the emission in the \ion{Ca}{2} H \& K line
cores, the Lithium abundance, and rotation. The new stellar parameters
yield improved values for the planetary mass and radius of $M_p =
1.198\pm0.053~M_\mathrm{Jup}$ and $R_p =
1.220^{+0.045}_{-0.042}~R_\mathrm{Jup}$, confirming that \mbox{TrES-2}
is the most massive among the currently known nearby ($d\lesssim 300$
pc) transiting hot Jupiters. The surface gravity of the planet, $\log
g_p = 3.299 \pm 0.016$, can be derived independently of the knowledge
of the stellar parameters (i.e., directly from observations), and with
a very high precision rivaling that of the best known double-lined
eclipsing binaries.

\end{abstract}





\section{Introduction}
\label{sec:introduction}

Our understanding of the structural and evolutionary properties of
close-in extrasolar planets (radius, mass, density) is continuously
improved by new detections of transiting planets. Fourteen such
systems are known to date\footnote{For a summary of their properties
see, for example,~\citeauthor{burrows07}~\citeyear{burrows07} or
obswww.unige.ch/$\sim$pont/TRANSITS.htm~.}. The accelerated rate at
which such systems have been discovered of late suggests the prospects
are bright for transit-search projects, as well as for the possibility
of critically testing physical models of hot Jupiters in the near
future based on statistically significant ensemble properties of
transiting planet systems (for a review
see~\citeauthor{charbonneau07}~\citeyear{charbonneau07}).

The accurate determination of the physical properties of transiting
exoplanets depends critically upon our knowledge of a number of basic
parameters of the parent stars. In particular, the mass and radius of
a planet, which are of fundamental importance for testing theoretical
predictions of planetary structure (e.g.,
~\citeauthor{guillot02}~\citeyear{guillot02};
~\citeauthor{bodenheimer03}~\citeyear{bodenheimer03};
~\citeauthor{baraffe03}~\citeyear{baraffe03};
~\citeauthor{burrows07}~\citeyear{burrows07}. 
For a review see for
example~\citeauthor{guillot05}~\citeyear{guillot05} and
~\citeauthor{burrows05}~\citeyear{burrows05}, and references therein),
depend rather directly on the mass and radius of the parent star,
placing strict demands on the accuracy of the latter. Evidence for
correlations between transiting planet properties and stellar
characteristics such as metallicity
(\citeauthor{guillot06}~\citeyear{guillot06};
~\citeauthor{burrows07}~\citeyear{burrows07}), and their implications
for competing giant planet formation models (e.g.,
~\citeauthor{ida04a}~\citeyear{ida04a},~\citeyear{ida04b};
~\citeauthor{kornet05}~\citeyear{kornet05};
~\citeauthor{alibert05}~\citeyear{alibert05};
~\citeauthor{boss00}~\citeyear{boss00},~\citeyear{boss02};
~\citeauthor{mayer04}~\citeyear{mayer04}), relies in turn on the
accurate determination of chemical abundances of the host
stars. Attempts to constrain the amount of mass loss experienced by
hot and very-hot Jupiters (\citeauthor{melo06}~\citeyear{melo06}), to
refine our knowledge of their relative frequencies
(e.g.,~\citeauthor{gaudi05}~\citeyear{gaudi05}) as well as to compare
observations with theoretical evaporation rates of insolated giant
planets (\citeauthor{lammer03}~\citeyear{lammer03};
~\citeauthor{baraffe04}~\citeyear{baraffe04};
~\citeauthor{lecavelier04}~\citeyear{lecavelier04};
~\citeauthor{lecavelier06}~\citeyear{lecavelier06}) are severely
affected by large uncertainties in the determination of stellar ages,
which can become pathological for field stars, depending on spectral
type.  Finally, the accurate determination of parent star parameters
requires particular attention in cases in which a direct distance
estimate to the system (trigonometric parallax) is unavailable. For
about 2/3 of the presently known nearby transiting systems ($d
\lesssim 300$ pc), and for over 3/4 of the full sample, such
measurements are not available at the present time, and will only be
made possible by future high-precision astrometric observatories, both
on the ground and in space
(e.g.,~\citeauthor{sozzetti05}~\citeyear{sozzetti05}, and references
therein).

Among the recently discovered transiting extrasolar planets, \mbox{TrES-2}
(\citeauthor{odonovan06}~\citeyear{odonovan06}) is the first detected
in the field of view of the {\it Kepler} mission
(\citeauthor{borucki03}~\citeyear{borucki03}). It has the largest
impact parameter, and is the most massive planet, among the currently
known nearby transiting systems
(\citeauthor{charbonneauetal07}~\citeyear{charbonneauetal07}).  In
this work we report on a detailed spectroscopic determination of the
properties of the parent star of \mbox{TrES-2}, including the effective
temperature and surface gravity, as well as the chemical abundances of
iron and lithium. We also measure the chromospheric activity and
provide constraints on the age of the system from this and other
indicators. Because of the importance of the effective temperature for
deriving other stellar characteristics, we have made an effort to
provide several external checks on its accuracy.  We then use these
properties along with other constraints from the light curve analysis
of Holman et al.~\citeyearpar{holman07b} to infer the mass and radius
of the star with realistic uncertainties. In particular, we show how
the use of the \emph{stellar density} obtained directly from the light
curve fit is in this case a much better proxy for luminosity than the
spectroscopic surface gravity, typically used in cases such as this in
which the parallax is unknown. Our new stellar parameters in turn lead
to improved values for the mass and radius of the planet over those
reported by O'Donovan et al.~\citeyearpar{odonovan06}. We conclude by
providing a summary of our results and by revisiting some of the
evidence connecting the properties of close-in extrasolar planets to
the characteristics of their parent stars.

\section{Observations}
\label{sec:observations}

The spectroscopic observations used here are the same as described
previously by O'Donovan et al.~\citeyearpar{odonovan06}. Briefly, they
consist of twelve echelle spectra obtained with the HIRES spectrograph
on the Keck~1 telescope (\citeauthor{vogt94}~\citeyear{vogt94}) during
the summer of 2006, with a nominal resolving power $R\simeq
71\,000$. Eleven of these spectra were obtained with an I$_2$ cell
placed in front of the slit to provide a wavelength fiducial for
high-precision velocity determinations (see,
e.g.,~\citeauthor{butler96}~\citeyear{butler96}), with typical
exposure times of 15 min resulting in an average $S/N\simeq 120$
pixel$^{-1}$.  One additional spectrum was obtained without the cell
for use as a template.  The effective wavelength coverage provided by
the three-CCD array of HIRES is $\sim$3200--8800\,\AA.  Additionally
we used 5 echelle spectra obtained with the Center for Astrophysics
(CfA) Digital Speedometers
(\citeauthor{latham92}~\citeyear{latham92}), which cover 45\,\AA\
centered at 5187\,\AA\ at a resolving power $R\simeq 35\,000$, and
have $S/N$ ratios ranging from 10 to 15 per resolution element.

\section{Atmospheric parameters}
\label{sec:atmospheric}

A detailed analysis of the template spectrum obtained with the Keck
telescope was carried out following the same procedures described in
detail by Sozzetti et
al. (\citeyear{sozzetti04},~\citeyear{sozzetti06}, and references
therein) in order to determine the effective temperature ($T_{\rm
eff}$), surface gravity ($\log g$), and iron abundance [Fe/H] of
\mbox{TrES-2}. A set of 30 relatively weak lines of \ion{Fe}{1} and 4
of \ion{Fe}{2} were selected, and equivalent widths (EWs) were
measured using the {\tt splot\/} task in IRAF\footnote{IRAF is
distributed by the National Optical Astronomy Observatories, operated
by the Association of Universities for Research in Astronomy, Inc.,
under contract with the National Science Foundation, USA.}. Metal
abundances are derived under the assumption of local thermodynamic
equilibrium (LTE), using the 2002 version of the spectral synthesis
code MOOG
(\citeauthor{sneden73}~\citeyear{sneden73})\footnote{http://verdi.as.utexas.edu/moog.html~.},
a grid of Kurucz ATLAS plane-parallel model stellar atmospheres
(\citeauthor{kurucz93}~\citeyear{kurucz93}), and imposing excitation
and ionization equilibrium. We obtained $T_{\rm eff} = 5850 \pm 50$~K,
$\log g = 4.4 \pm 0.1$, $\xi_t = 1.00 \pm 0.05$ km s$^{-1}$, and
[Fe/H] = $-0.15\pm 0.10$.  The uncertainties in the first three
parameters were estimated following the prescriptions of Neuforge \&
Magain~\citeyearpar{neuforge97} and Gonzalez \&
Vanture~\citeyearpar{gonzvant98}, and were rounded off to the nearest
25~K in $T_\mathrm{eff}$, 0.1 dex in $\log g$, and 0.05 km s$^{-1}$ in
$\xi_t$. For [Fe/H] the uncertainty given corresponds to the scatter
obtained from the \ion{Fe}{1} lines rather than the formal error of
the mean, since we consider the latter to be unrealistically small in
this case.  No significant departures from LTE are expected for a star
with the temperature and metallicity of \mbox{TrES-2}
(e.g.,~\citeauthor{yong04}~\citeyear{yong04}), so for the purpose of
this study we have not included non-LTE effects in our spectroscopic
analysis. We also quantified the sensitivity of our iron abundance
determination to variations of $\pm1\sigma$ with respect to the
nominal $T_\mathrm{eff}$, $\log g$, and $\xi_t$ values, and found
changes in [Fe/H] of at most 0.06 dex, below the adopted uncertainty
of 0.1 dex.  Finally, we determined also the projected rotational
velocity as $v\sin i = 2.0\pm 1.0$ km s$^{-1}$, based on the synthesis
of a set of unblended \ion{Fe}{1} lines, following
Gonzalez~\citeyearpar{gonzalez98}.

The new values for the stellar parameters are consistent with the
\ion{G0}{5} spectral type implied by the colors
(\citeauthor{odonovan06}~\citeyear{odonovan06}), and are in generally
good agreement with those presented by those authors.  We note,
however, that our $T_\mathrm{eff}$ value is somewhat lower, possibly
due to the fact that O'Donovan et al.~\citeyearpar{odonovan06} assumed
solar metallicity in their study, whereas our analysis indicates a
slightly metal-deficient composition.

\subsection{Consistency checks on the $T_\mathrm{eff}$ estimate}

Given the importance of the temperature determination for establishing
the absolute mass and radius of the parent star of \mbox{TrES-2}, we present
in this section a number of other consistency checks that illustrate
the reliability and accuracy of our estimate above.

\subsubsection{Estimate from the CfA spectra}
\label{sec:cfa}

Cross-correlation of our CfA spectra against a large library of
synthetic templates in the manner described by O'Donovan et
al.~\citeyearpar{odonovan06} provides an independent estimate of the
photospheric properties of the star. By testing all combinations of
the four main parameters of these templates ($T_\mathrm{eff}$, $\log
g$, metallicity [m/H], and $v \sin i$) we seek to maximize the
correlation averaged over the 5 available spectra. In principle this
allows the determination of the four parameters, although in practice
the narrow wavelength coverage of only 45\,\AA\ results in strong
correlations between some of those properties. We therefore determined
the first two of these quantities along with $v \sin i$ for fixed
metallicities of [m/H] = 0.0 (solar) and [m/H] $= -0.5$, which bracket
our determination in \S\ref{sec:atmospheric}, and then interpolated to
[m/H] $= -0.15$.  The results are $T_\mathrm{eff} = 5790 \pm 100$~K,
$\log g = 4.3 \pm 0.2$, and $v \sin i = 1.0 \pm 2.0$~km s$^{-1}$,
which supersede the values given by O'Donovan et
al.~\citeyearpar{odonovan06} that were based on a preliminary
analysis. The temperature is only 60~K lower than our determination in
\S\ref{sec:atmospheric}, well within the errors. There is also
excellent agreement in $\log g$ and $v \sin i$.
	
\subsubsection{Spectral line-depth ratios}

Among the many diagnostics available for effective temperature
estimation, the ratio of the depths of two spectral lines having
different sensitivity to temperature is among the most powerful.
Indeed, the line-depth ratio (LDR) technique allows the measurement of
temperature {\it differences} of the order of a few Kelvin in
favorable cases (\citeauthor{gray91}~\citeyear{gray91};
~\citeauthor{gray94}~\citeyear{gray94};
~\citeauthor{strassmeier00}~\citeyear{strassmeier00};
~\citeauthor{gray01}~\citeyear{gray01};
~\citeauthor{catalano02}~\citeyear{catalano02};
~\citeauthor{caccin02}~\citeyear{caccin02};
~\citeauthor{kovtyukh03}~\citeyear{kovtyukh03};
~\citeauthor{biazzo07}~\citeyear{biazzo07}), a much higher accuracy
than currently possible with other methods that seek to determine
temperatures on an absolute scale. The latter typically have
uncertainties of at least 50--100~K. Absolute temperature
determinations with the LDR method still require translation to an
absolute scale, usually through color/temperature calibrations since
the LDR measurements are typically calibrated first against color
indices, which are directly observable.

In order to provide a further check of our spectroscopic
$T_\mathrm{eff}$ determination, we selected in our Keck template
spectrum 9 pairs of temperature-sensitive lines from the list of
Biazzo et al.~\citeyearpar{biazzo07} in the red part of the spectrum
($\lambda = 6190$--6280\,\AA). These authors provide calibrations
between LDRs and temperature for a range of rotational velocities
(those for zero rotation were adopted here), which include corrections
for surface gravity and are valid in the range 3800~K~$\lesssim
T_\mathrm{eff}\lesssim 6000$~K.  Metallicity effects are negligible
for stars near the solar abundance
(\citeauthor{gray94}~\citeyear{gray94};~\citeauthor{biazzo07}~\citeyear{biazzo07}). The
absolute temperatures derived in this way rely implicitly on an
intermediate calibration between the $B-V$ color and $T_\mathrm{eff}$
adopted from Gray~\citeyearpar{gray05}. In order to provide
consistency with the calibrations we use below, we have converted each
of our 9 temperature estimates back into $B-V$ and then into
temperature again adopting the calibrations by Ram\'\i rez \&
Mel\'endez~\citeyearpar{ramirez05}, which have been compared carefully
against absolute temperature determinations using the Infrared Flux
Method. The average temperature we obtain in this way is $5780 \pm
50$~K, in agreement with our estimates above.

\subsubsection{H$_\alpha$ line profile}

\begin{figure}
\plotone{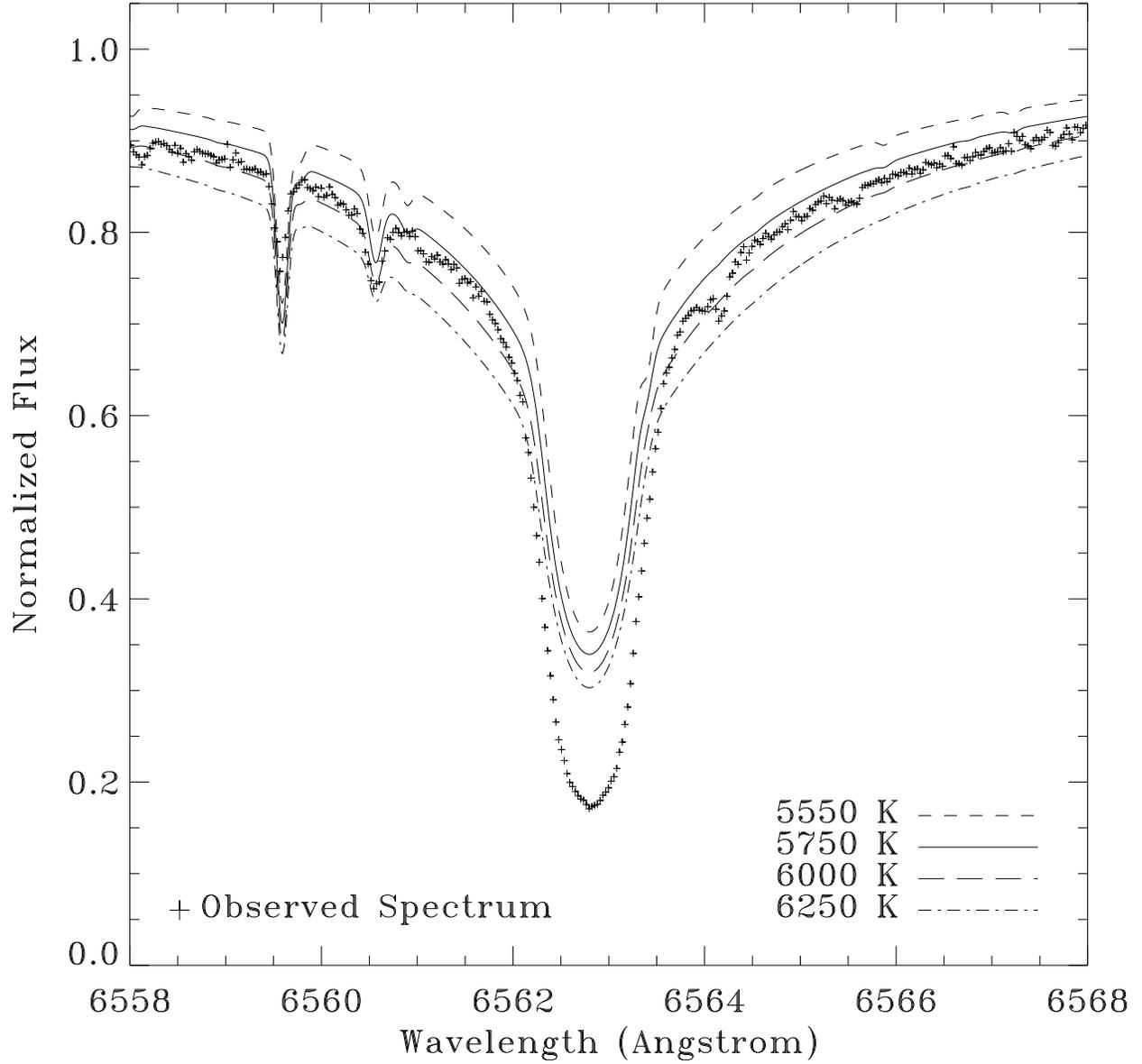}
\caption{Observed H$_\alpha$ profile in the Keck template spectrum of
\mbox{TrES-2} compared with four synthetic spectra with [m/H] = 0.0, $\log g
= 4.5$, and effective temperatures of 5500, 5750, 6000, and 6250 K,
respectively}. \label{halpha}
\end{figure}

The strong sensitivity of the wings of the H$_\alpha$ line profile to
temperature variations, as well as the relatively weak sensitivity to
changes in surface gravity and metal abundance, make this feature a
very useful temperature indicator for solar-type dwarfs
(e.g.,~\citeauthor{fuhrmann93}~\citeyear{fuhrmann93};
~\citeauthor{barklem02}~\citeyear{barklem02}). The core of the line,
however, is formed higher up in the atmosphere under conditions that
violate LTE, so is not useful here.  As an additional consistency
check on $T_\mathrm{eff}$ we therefore compared the H$_\alpha$ line
profile outside of the core in our Keck template spectrum against
synthetic profiles for solar-metallicity dwarfs ([m/H] = 0.0, $\log g
= 4.5$) from the Kurucz database.  In Figure~\ref{halpha} we show the
normalized flux in a 10~\AA\ region centered on H$_\alpha$, and four
calculated profiles for different values of $T_\mathrm{eff}$.
Temperatures significantly hotter than 6000~K or significantly cooler
than 5750~K appear inconsistent with the observed profile, and suggest
the optimal value is somewhere in between.  Because of the
difficulties in the placement of the continuum for such a broad line
in an echelle spectrum, we view this comparison only as a rough
check. Nevertheless, it agrees once again with our previous estimates.

\subsubsection{Photometric estimates}
\label{sec:photometry}

Photometric measurements for \mbox{TrES-2} are available in the
Johnson, Cousins, {\it Tycho\/}, and 2MASS systems, as listed by
O'Donovan et al.~\citeyearpar{odonovan06}.  Based on these data and
assuming zero reddening we have computed 7 different color indices
(not all completely independent of each other) and applied the
calibrations of Ram\'\i rez \& Mel\'endez~\citeyearpar{ramirez05} for
a fixed metallicity of [Fe/H] $= -0.15$ to obtain effective
temperatures.  There is excellent agreement between these estimates,
and the average $T_\mathrm{eff}$ is $5680 \pm 50$~K. This is 170~K
cooler than our spectroscopic determination in
\S\ref{sec:atmospheric}, a difference that is significantly larger
than allowed by the combined uncertainties.  However, in view of the
distance to the object of $\sim$230~pc
(\citeauthor{odonovan06}~\citeyear{odonovan06}; see also below), a
small amount of reddening would not be entirely unexpected, and was
also suspected by those authors. We find that if we applied a
correction to the individual color indices corresponding to $E(B-V)
\approx 0.04$~mag, the average temperature would agree exactly with
our spectroscopic value. Indirect support for this amount of
interstellar material is provided by the reddening maps of Burstein \&
Heiles~\citeyearpar{burstein82} and Schlegel et
al.~\citeyearpar{schlegel98}, which indicate total $E(B-V)$ values
along the line of sight to \mbox{TrES-2} of $\sim$0.05~mag and
$\sim$0.07~mag, respectively. For the distance of \mbox{TrES-2} and
using the Drimmel \& Spergel~\citeyearpar{drimmel01} model of Galactic
dust distribution these numbers are reduced to $E(B-V) \approx 0.02$
and 0.03~mag, respectively. Further support is given by the comparison
between the observed colors and those predicted from stellar evolution
models described below, which gives $E(B-V) \approx 0.03 \pm
0.02$. 

From the good agreement between our spectroscopic temperature
determination in \S\ref{sec:atmospheric} and the estimates from the
CfA spectra and the LDRs we conclude that the $T_\mathrm{eff}$ of
\mbox{TrES-2} is accurately established and we make use of it in
\S\ref{sec:physics} to infer the mass and radius of the star.

\section{Constraints on the stellar age}

Age determination for individual stars in the field is a difficult
task. A variety of indicators is available, such as H$_\alpha$
emission, X-ray activity, lithium abundance, \ion{Ca}{2} H \& K
emission, asteroseismology, rotation, and Galactic space motion, some
more constraining than others depending on the particular case. While
the relative agreement among multiple methods allows in principle for
fairly reliable dating of stars with ages comparable to the age of the
Hyades or younger, constraints on ages for individual stars with
$t\gtrsim 1-2$ Gyr are usually quite weak.  Here we have used the
\ion{Ca}{2} activity indicator and the lithium abundance as measured
in our HIRES spectra of \mbox{TrES-2} in an attempt to inform the theoretical
models used for the determination of the stellar mass and radius (see
next Section).

\begin{figure}
\centering
$\begin{array}{c}
\includegraphics[width=0.60\textwidth]{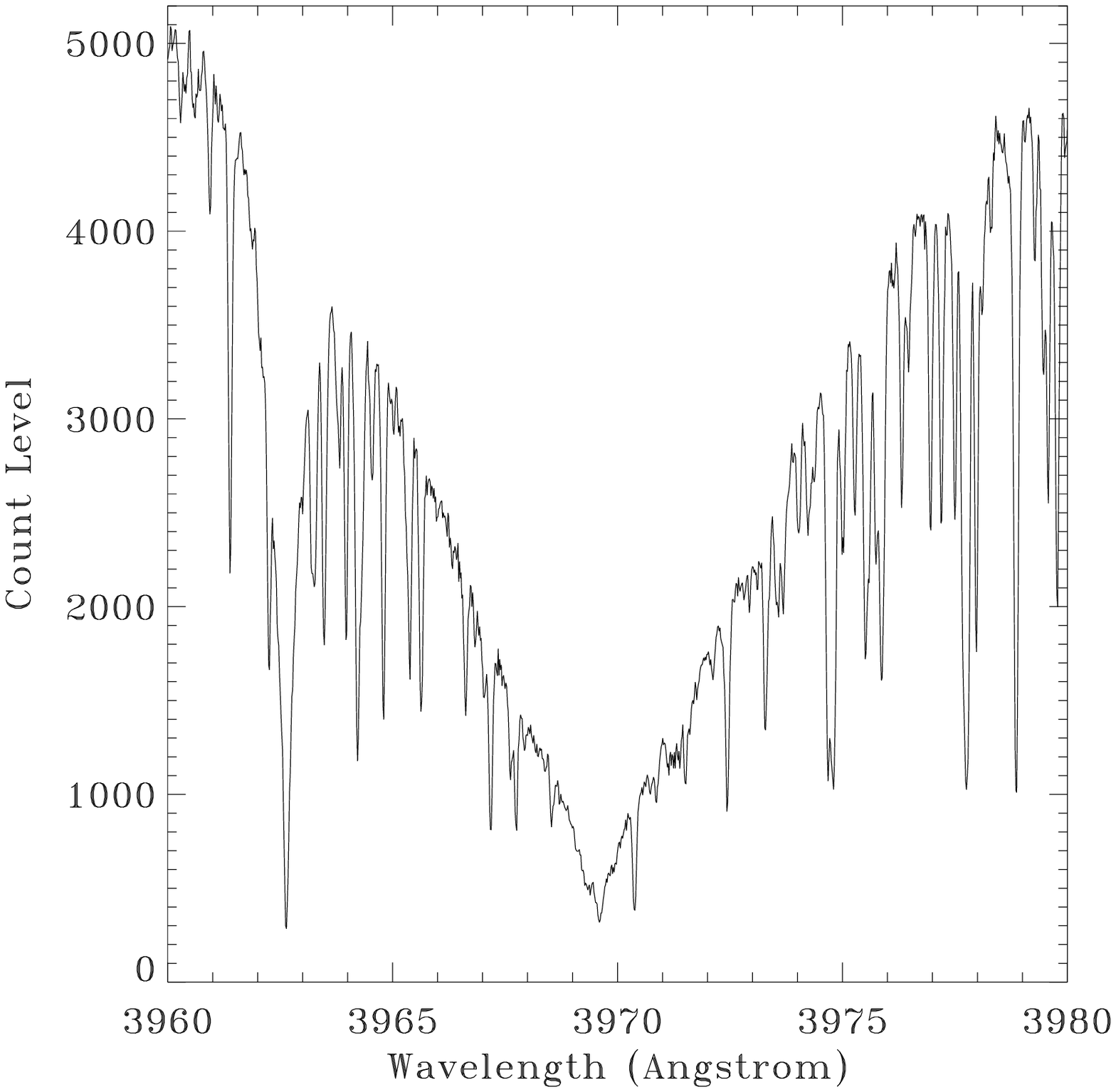} \\
\includegraphics[width=0.70\textwidth]{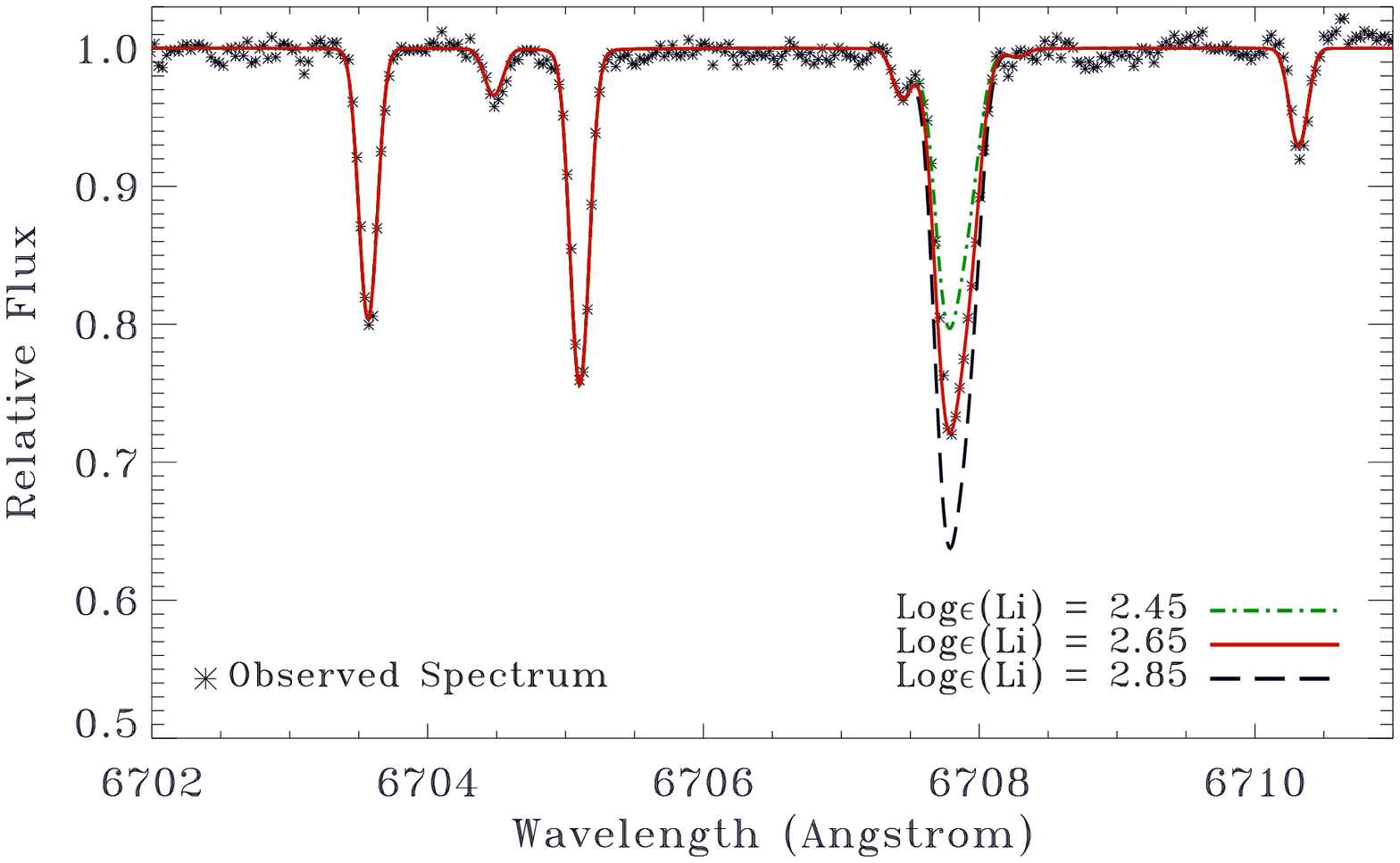} \\
\end{array}$
\caption{Top: a 10 \AA\ region of the Keck template spectrum of \mbox{TrES-2}
centered on the \ion{Ca}{2} H line.  Bottom: a portion of the same
spectrum containing the \ion{Li}{1} line at 6707.8 \AA\ (filled dots),
compared to three synthetic profiles (lines of various colors and
styles), each differing only in the lithium abundance
assumed. \label{cali}}
\end{figure}

In the top panel of Figure~\ref{cali} we show a region of the HIRES
template spectrum centered on the \ion{Ca}{2} H line. No significant
emission feature is present, and the same is true for the other Keck
spectra we collected using the I$_2$ cell (in which the iodine lines
do not interfere because there are none shortward of $\sim$5000\,\AA).
Following the procedure outlined in Sozzetti et
al.~\citeyearpar{sozzetti04}, we have measured the chromospheric activity 
index S (e.g., Duncan et al. 1991) from the \ion{Ca}{2} H and K lines in our spectra, 
and then converted it into the chromospheric emission ratio 
$\log R^\prime_{HK}$, corrected for the photospheric contribution. 
For \mbox{TrES-2}, the Mount Wilson index, averaged over all our spectra, is 
$\langle S\rangle = 0.13$, and the resulting $\langle\log R^\prime_{HK}\rangle = -5.16\pm0.15$ 
(with formal uncertainties calculated from the scatter of individual measurements), 
suggesting a quite inactive star. The resulting chromospheric age estimate, using the
relations summarized in Wright et al.~\citeyearpar{wright04}, is $t =
8.32\pm1.07$ Gyr.

We point out, however, that this estimate should not be taken at face
value for a number of reasons. Firstly, it has been shown (e.g.,
~\citeauthor{pace04}~\citeyear{pace04}) that chromospheric activity
can only be considered a reliable age estimator up to $t\simeq 2$
Gyr, and that for stars showing low activity levels the ages derived
are only lower limits.  Secondly, because stars have activity cycles
like the Sun, the proper measure of the chromospheric flux to use in
determining the age of an individual star is an average over the
entire magnetic cycle rather than a quasi-instantaneous value such as
is available to us, to avoid the possibility of finding a star in a
Maunder minimum phase
(\citeauthor{henry96}~\citeyear{henry96};~\citeauthor{wright04}~\citeyear{wright04}).
Finally, there are hints (\citeauthor{song04}~\citeyear{song04}) that
chromospheric age estimates tend to be systematically older than those
derived with other methods, suggesting perhaps the need for
re-calibration of the \ion{Ca}{2} activity-based ages. Therefore, all
we can claim here is that the lower limit for the chromospheric age of
\mbox{TrES-2} is a few Gyr.

\begin{figure}
\plotone{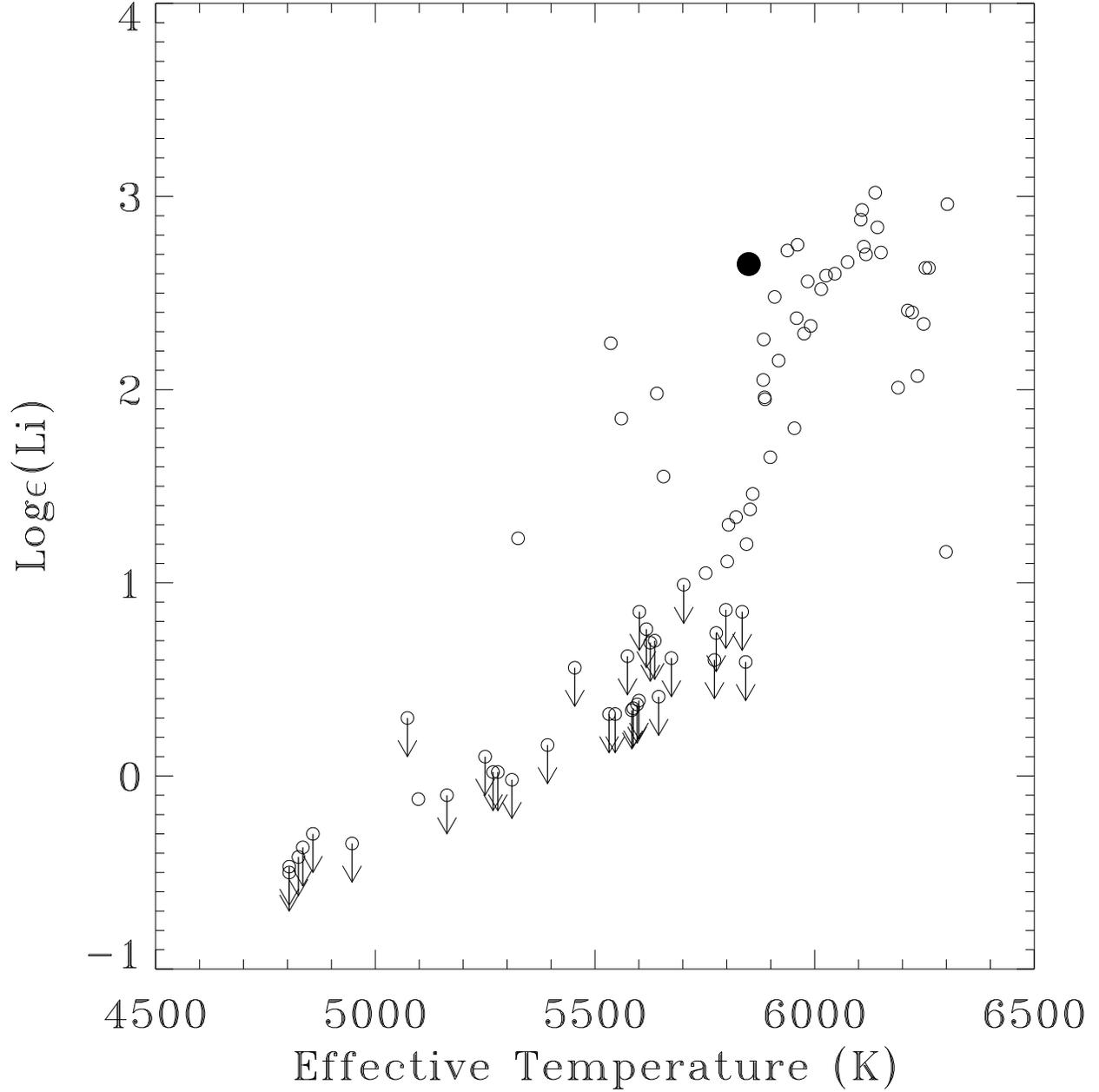}
\caption{Lithium abundance $\log\epsilon$ (Li) as a function of
effective temperature $T_\mathrm{eff}$ for \mbox{TrES-2} (filled circle) and
a sample of planet hosts (open circles, data from
\citeauthor{israelian04}~\citeyear{israelian04}). 
Arrows indicate that only upper limits on $\log\epsilon$
(Li) are available.} \label{licomp}
\end{figure}

\mbox{TrES-2} displays a significant Li~$\lambda$6707.8 absorption feature.
We have carried out a detailed spectral synthesis of a 10~\AA\ region
of the Keck template spectrum centered on this line, using the
atmospheric parameters derived from the Fe-line analysis and the line
list of Reddy et al.~\citeyearpar{reddy02}.  In the bottom panel of
Figure~\ref{cali} we show the comparison of the spectrum of \mbox{TrES-2}
with three synthetic spectra, each differing only in the Li abundance
assumed. The best-fit model results in an abundance of
$\log\epsilon{\rm (Li)} = 2.65$. As shown in Figure~\ref{licomp}, the
Li abundance we obtain does not appear out of the ordinary in relation
to those of other planet host stars of similar temperature, as
measured by Israelian et al.~\citeyearpar{israelian04} (once typical uncertainties 
of the order of 50-100 K on the $T_\mathrm{eff}$ determinations are considered). 
By comparison with Li abundance curves as a function of effective temperature for
clusters of different ages
(\citeauthor{sestito05}~\citeyear{sestito05}) we infer for \mbox{TrES-2} an
age of about 1--2 Gyr. This value would point to a somewhat younger age than 
the one inferred from the \ion{Ca}{2} measurements. However, it is not uncommon to observe a large
spread in Li abundance among stars in the same cluster that appear
otherwise identical (\citeauthor{randich06}~\citeyear{randich06}), so
also in this case it seems safer to simply report a lower limit to the
age of \mbox{TrES-2} of 1--2 Gyr
\footnote{However, we point out how one can speculate on the possibility 
that self-enrichment (see, e.g., Gonzalez 2006, and references therein, 
for a review of the issue), rather than systematics or uncertainties in 
the calibration of activity-age relations, could be a factor to consider for 
TrES-2. If the star has witnessed recent events of accretion of planetary material, 
this could explain both the somewhat higher than usual Lithium abundance with respect to other 
planet hosts of the same $T_\mathrm{eff}$, as well as the apparent discrepancy between 
the youth indicators. The measurement of statistically significant trends of element 
abundance with condensation temperature $T_c$ (e.g., Sozzetti et al. 2006; Gonzalez 2005, 2006, 
and references therein) or detection of the $^6$Li isotope (e.g., Israelian et al. 2001, 2003; 
Gonzalez 2006, and references therein) in the atmosphere of TrES-2 would be strong 
evidence in support of the self-enrichment scenario for this star. Further spectroscopic 
measurements of TrES-2 are thus clearly encouraged}.

Finally, another argument for the star not being particularly young is
given by the small projected rotational velocity we measure here (see
\S\ref{sec:atmospheric} and \S\ref{sec:cfa}). This would indicate once
again an age of $t\gtrsim 1-2$ Gyr
(\citeauthor{bouvier97}~\citeyear{bouvier97};
~\citeauthor{pace04}~\citeyear{pace04}).

\section{Stellar mass and radius}
\label{sec:physics}

A common procedure for deriving the absolute mass and radius of planet
host stars, needed to infer those of the transiting object, is to
compare the measured stellar properties such as temperature and
luminosity with stellar evolution models in the H-R diagram, or in
some equivalent parameter space.  Because the distance to \mbox{TrES-2} is
not precisely known (it was not observed during the {\it Hipparcos\/}
mission), we do not have direct access to its luminosity. An
alternative measure of intrinsic brightness (or evolution) that has
been used in the past is the spectroscopically determined value of
$\log g$ (see, e.g.,
~\citeauthor{konacki03}~\citeyear{konacki03},~\citeyear{konacki04},~\citeyear{konacki05};
~\citeauthor{pont04}~\citeyear{pont04};
~\citeauthor{bouchy05}~\citeyear{bouchy05};
~\citeauthor{santos06}~\citeyear{santos06}). Surface gravities are
typically very difficult to determine accurately in this way, and as a
result the constraint on the stellar radius is relatively weak. Here
we explore in detail the possibility of using other information
available in transiting systems such as \mbox{TrES-2}, that provide much
tighter constraints on the luminosity, as also noted recently by Pont
et al.~\citeyearpar{pont07}.  We focus in particular on the quantities
obtainable by fitting the transit light curves.  The three main
adjustable parameters (see,
e.g.,~\citeauthor{mandel02}~\citeyear{mandel02}) are often taken to be
the relative radius of the planet ($R_p/R_\star$), the impact
parameter ($b \equiv a \cos i/R_\star$), and the normalized separation
between the star and the planet ($a/R_\star$), where $a$ is the
semimajor axis of the relative orbit and $i$ is the inclination to the
line of sight. These are largely independent of the stellar
properties, except for a weak dependence on the limb-darkening
coefficients (a second-order effect) that are typically a function of
effective temperature, surface gravity, and composition\footnote{In
some cases even this weak dependence can be avoided altogether by
fitting for the limb-darkening coefficients simultaneously with the
other three parameters of the transit light curve.}. One of these,
$a/R_\star$, contains information intrinsic to the star: using
Kepler's Third Law (as revised by Newton) it can be shown that
\begin{equation}
\label{eq:density}
{M_\star \over R_{\star}^3} = {4 \pi^2\over G P^2}\left({a\over
R_\star}\right)^3 - {M_p \over R_{\star}^3}~.
\end{equation}
(see also~\citeauthor{seager03}~\citeyear{seager03}) where all
quantities are expressed in {\it cgs\/} units and $G$ is the Newtonian
gravitational constant.  The left-hand side corresponds essentially to
the stellar density, $\rho_\star$.  Note that the first term on the
right is entirely determined from measurable quantities: the orbital
period ($P = 2.47063 \pm 0.00001$
days;~\citeauthor{odonovan06}~\citeyear{odonovan06}), and $a/R_\star$
from the light curve fit.  The second term on the right, on the other
hand, involves the planetary mass (which is unknown until the stellar
mass is determined) as well as $R_\star$ (also unknown). However, the
size of this second term is typically two to three orders of magnitude
smaller than the first for most transiting exoplanets including
\mbox{TrES-2}, so it can safely be ignored for all practical
purposes. Thus the density of the star is determined \emph{directly}
from the observations, with no additional assumptions. In this
particular case the accuracy of the $a/R_\star$ determination ($7.63
\pm 0.12$; ~\citeauthor{holman07b}~\citeyear{holman07b}) is very high:
the uncertainty is only 1.6\%.  The stellar density is a sensitive
measure of evolution or luminosity, and as such it provides a very
useful constraint on the size of the star. We use it along with the
effective temperature and the measured metallicity to establish the
absolute mass and radius of \mbox{TrES-2}.

We proceed by comparing $\rho_\star$ (or a closely related quantity;
see below) and $T_\mathrm{eff}$ with current stellar evolution models
from the Yonsei-Yale series by Yi et al.~\citeyearpar{yi01} (see
also~\citeauthor{demarque04}~\citeyear{demarque04}). We explored the agreement 
with model isochrones calculated over a wide a range of uniformly spaced ages 
(0.1--9 Gyr) spanning the full range of metallicities allowed by our
spectroscopic determination ([Fe/H] $= -0.15 \pm 0.10$). Along each
isochrone we computed the theoretical stellar properties using a fine
step in mass, and at each of these points we compared those properties
with the observations, and recorded all cases yielding a match within
the observational errors. In this way we established the range of
permitted values of the stellar mass and radius. All these matching 
models were assigned the same likelihood for this application. In practice, 
we have chosen to compare the models with the observations directly in the
observational plane ($a/R_\star$ versus $T_\mathrm{eff}$, rather than
$\rho_\star$ versus $T_\mathrm{eff}$). Therefore, instead of computing
the stellar density along the isochrones and comparing it with the
measured value of $\rho_\star$, we computed the theoretical values of
$a/R_\star$ (which is essentially the cube root of the density) from
an expression obtained by rearranging eq.~\ref{eq:density},
\begin{equation}
\label{eq:aR}
{a\over R_\star} = \left({G\over 4\pi^2}\right)^{1/3} {P^{2/3}\over
R_\star} \left(M_\star + M_p\right)^{1/3}~,
\end{equation}
and compared them with the value resulting from the light-curve fit,
$a/R_\star = 7.63 \pm 0.12$
(\citeauthor{holman07b}~\citeyear{holman07b}).  As discussed above,
we have ignored here the small contribution from the mass of the
planet, $M_p$.\footnote{If the need ever arose (for example, for much
more massive planets), it would be trivial to account for this small
correction term by simply using a rough estimate of the planet mass in
eq.~\ref{eq:aR} to compute the predicted values of $a/R_\star$.}

The best match with the models (which produces virtually perfect
agreement with the measured parameters $a/R_\star$ and
$T_\mathrm{eff}$) is achieved for a stellar mass of $M_\star = 0.980
\pm 0.062~M_\sun$, a radius of $R_\star =
1.000^{+0.036}_{-0.033}~R_\sun$, and an age of $5.1^{+2.7}_{-2.3}$
Gyr.  The uncertainties listed reflect the spread allowed by the
observational errors in $T_\mathrm{eff}$, $a/R_\star$, and [Fe/H], and
exclude any systematics in the models themselves, which are difficult
to quantify. The surface gravity of the star for the best fit is $\log
g = 4.429^{+0.021}_{-0.023}$, in excellent agreement with our
spectroscopically measured value in \S\ref{sec:atmospheric}, and the
corresponding metal abundance for this best fit is [Fe/H] $= -0.14$,
also virtually the same as the measured value. The models indicate for
the star an age similar to the Sun, which is entirely consistent with
the lower limits discussed earlier based on the activity indicators
and Li.

The absolute visual magnitude predicted by the models for the adopted
mass and age and our measured composition is $M_V = 4.77 \pm 0.09$,
and the color expected is $B-V = 0.591 \pm 0.014$. The latter,
compared with the measured index of $0.619 \pm 0.009$ (O'Donovan et
al.\ 2006) suggests a small amount of reddening of $E(B-V) = 0.03 \pm
0.02$ mag, not inconsistent with several other estimates discussed in
\S\ref{sec:photometry}. This corresponds to a visual extinction $A_V
\sim 0.1$ mag. Accounting for this we derive a distance to
\mbox{TrES-2} of 220 pc with an estimated uncertainty of 10 pc. With
the mean radial velocity of $RV = -0.56 \pm 0.11$~km~s$^{-1}$ as
reported by O'Donovan et al.\ (2006), and UCAC2 
(\citeauthor{zacharias04}~\citeyear{zacharias04}) proper motion
components [$\mu_\alpha$, $\mu_\delta$] = [4.45, $-$3.40]
mas~yr$^{-1}$, the Galactic space motion vector of the
star is [$U$, $V$, $W$] = [$-$1.81, +0.88, $-$5.51] km~s$^{-1}$ (where 
$U$ is taken to be positive toward the Galactic anti-center).  We
collect these results along with other properties derived previously
in Table~\ref{tab:star}.

Our stellar mass is $\sim$10\% smaller than the value of $M_\star =
1.08$~$M_\sun$ adopted by O'Donovan et al.~\citeyearpar{odonovan06},
which is in part due to our using a lower heavy element abundance,
derived from our detailed chemical analysis, as opposed to adopting
the solar composition. The stellar radius, on the other hand, is the
same.

\begin{figure}
\centering
\includegraphics[width=0.60\textwidth]{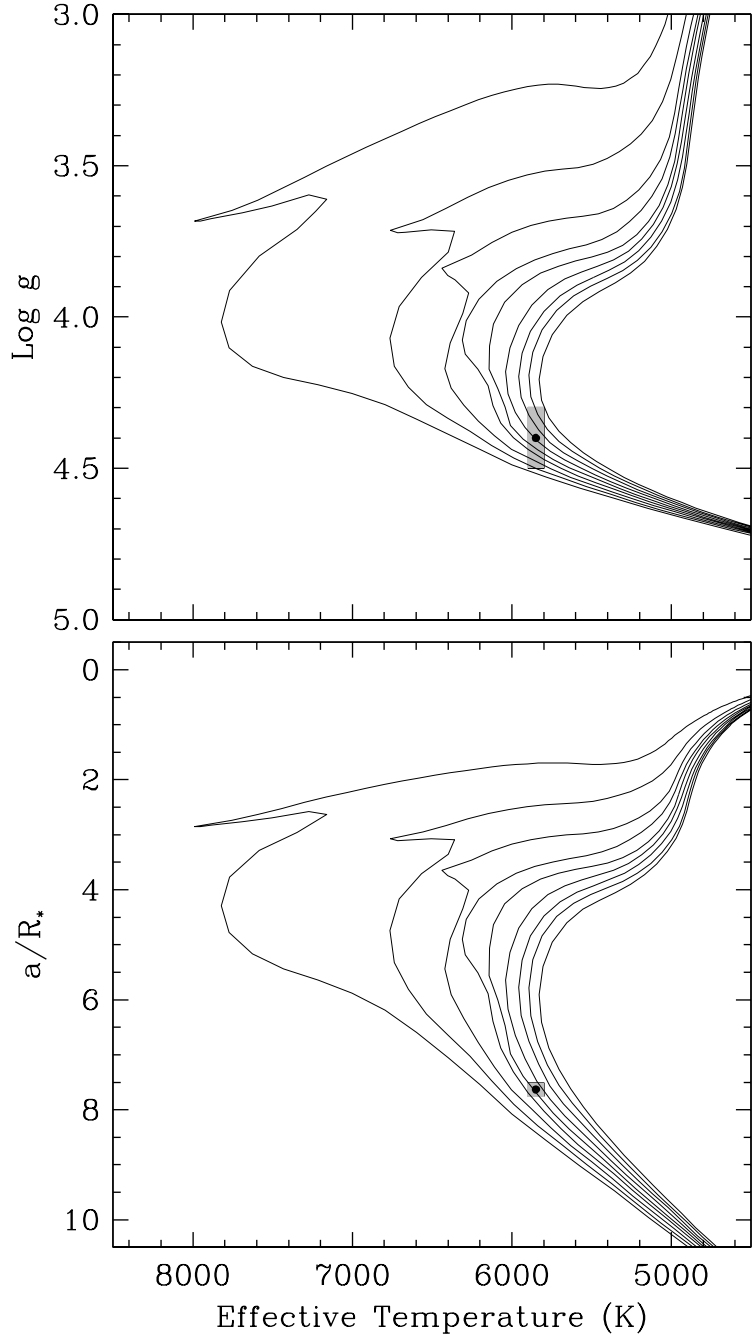}
\caption{Model isochrones from the Yonsei-Yale series by Yi et
al.~\citeyearpar{yi01} and Demarque et al.~\citeyearpar{demarque04},
corresponding to ages of 1--9 Gyr (left to right), for the measured
composition of [Fe/H] $= -0.15$, shown with the observational
constraints. {\it Top:} The measurement on the vertical axis is the
spectroscopically determined value of $\log g$, which provides only a
weak handle on the stellar radius and no useful constraint on
age. {\it Bottom:} The use of the photometrically determined value of
$a/R_\star$ from the light curve analysis of Holman et
al.~\citeyearpar{holman07b} instead of surface gravity provides a much
stronger constraint on the age and radius (by about a factor of 5).
\label{fig:iso}}
\end{figure}

The fairly tight constraints we have obtained on the radius of the
star (errors less than 4\%) are the result of using the information on
$a/R_\star$ derived from the light-curve fit of Holman et
al.~\citeyearpar{holman07b}.  Had we used the spectroscopically
determined surface gravity instead, we estimate the constraint would
have been some 5 times weaker. This is illustrated in
Figure~\ref{fig:iso}, where the top panel shows Yonsei-Yale isochrones
for ages of 1~Gyr to 9~Gyr and a heavy element abundance equal to the
measured value of [Fe/H].  The shaded error box represents the
uncertainties in the measured $\log g$ and $T_\mathrm{eff}$ of
\mbox{TrES-2}. As seen, the error in gravity is so large as to span the full
range of ages shown here, thus providing essentially no useful
constraint on age and a weak one on $R_\star$.  In the lower panel we
have re-mapped the vertical axis to $a/R_\star$ (using the measured
orbital period $P$; see eq.~\ref{eq:aR}). The error box in this case
is significantly smaller, making $a/R_\star$ a much better measure of
the luminosity than surface gravity. We propose that the same
procedure should be used in other transiting planets in which the
quality of the light curves is sufficient to provide a superior
constraint compared to surface gravity. Depending on the case, the
accuracy of $a/R_\star$ could be high enough that it may even compare
favorably with the constraint afforded by a direct knowledge of the
parallax.
	
\section{Revised planetary parameters}

The improved knowledge of the mass and radius of the parent star has a
direct impact on the accuracy of the planetary parameters of \mbox{TrES-2}.
We have combined the stellar properties in Table~\ref{tab:star} with
the mass function from the spectroscopic orbit of O'Donovan et
al.~\citeyearpar{odonovan06}, $M_p \sin i = 1.206 \pm 0.016 \left[
(M_\star + M_p)/M_\sun \right]^{2/3} M_\mathrm{Jup}$, and the fitted
light-curve parameters from the new photometric analysis of Holman et
al.~\citeyearpar{holman07b}, which are $R_p/R_\star = 0.1253 \pm
0.0010$, $b = 0.8540 \pm 0.0062$, and $a/R_\star = 7.63 \pm 0.12$. We
obtain for the planet $M_p = 1.198\pm0.053~M_\mathrm{Jup}$ and $R_p =
1.220^{+0.045}_{-0.042}~R_\mathrm{Jup}.$\footnote{The equatorial
radius adopted for Jupiter is 71\,492 km.} These are some 6\% and 1\%
smaller than the values reported by O'Donovan et
al.~\citeyearpar{odonovan06}, respectively.

An important but generally overlooked property of the spectroscopic
and photometric solutions for transiting planets is the fact that the
surface gravity of the planet is directly measurable from the
observations, with no need to know the mass or radius of the parent
star (see, e.g.,~\citeauthor{winn07b}~\citeyear{winn07b}). The same
was pointed out by ~\citeauthor{beatty07}~(\citeyear{beatty07}) in the
context of mass and radius determinations for small stars in 
single-lined eclipsing binaries, and also by Southworth et al.\ (2004, 2007).  
This derives from the quadratic relation between $M_p$ and $R_p$ that 
can easily be obtained from the definition of the spectroscopic mass 
function and Kepler's Third Law:
\begin{equation}
\label{eq:planetmass}
M_p = {2\pi\over G P} {K_\star \sqrt{1-e^2} \over
\sqrt{1-[b/(a/R_\star)]^2}} \left({a/R_\star \over
R_p/R_\star}\right)^2 R_p^2~.
\end{equation}
In this expression $K_\star$ represents the velocity semi-amplitude of
the star in response to the pull from the planet ($K_\star = 181.3 \pm
2.6$ m~s$^{-1}$ for
\mbox{TrES-2};~\citeauthor{odonovan06}~\citeyear{odonovan06}), and $e$ the
eccentricity of the orbit, which is usually found to be very close to
zero for transiting planets. The quantities $b$, $a/R_\star$, and
$R_p/R_\star$ are obtained directly from the light curve analysis,
often with very high precision as in our case.  The surface gravity of
the planet follows immediately as:
\begin{equation}
\label{eq:gravity}
\log g_p  = -2.1383 - \log P + \log K_\star - 
{1\over 2}\log \left(1-\left[{b \over a/R_\star}\right]^2\right) + 2
\log\left({a/R_\star \over R_p/R_\star}\right) + {1\over 2}\log\left(1-e^2\right)~.
\end{equation}
The numerical constant is such that the gravity is in {\it cgs\/}
units when $P$ and $K_\star$ are expressed in their usual units of
days and m~s$^{-1}$. For \mbox{TrES-2} we obtain $\log g_p = 3.299 \pm
0.016$, in which the uncertainty includes all contributions from
measured quantities. We call the reader's attention to the very high
precision of this determination, which rivals that of the best-known
double-lined eclipsing binaries (see,
e.g.,~\citeauthor{andersen91}~\citeyear{andersen91}).  While the
planetary masses and radii in transiting systems have typically been
the main focus of investigators in this field, and with good reason,
those quantities depend critically on the mass and radius of the
parent star, which are often the weak link in the chain and usually
rely on stellar evolution models. Surface gravities are much closer to
the observations, are model-independent, and can often be obtained
with very high precision as in the case of \mbox{TrES-2}.  Accurately
determined surface gravities of planets are potentially important for
constraining theoretical calculations of the spectra of extrasolar
planets. These have now begun to be tested through infrared photometry
of the secondary eclipses as well as transmission spectroscopy in
several cases (see for
example~\citeauthor{charbonneau07}~\citeyear{charbonneau07}, and
references therein).
	
\section{Summary and discussion}

Our high-resolution, high-$S/N$ spectra from Keck/HIRES have allowed
us to derive new and accurate values of the stellar atmospheric
parameters of the parent star of the transiting planet \mbox{TrES-2},
principally the effective temperature and metallicity, which have in
turn yielded improved parameters for the star. The G0V main-sequence
dwarf appears to have a metal abundance very similar to the average of
the solar neighborhood ([Fe/H]$\simeq -0.1$, see for
example~\citeauthor{nordstrom04}~\citeyear{nordstrom04}), making it
nominally the most metal-deficient case in the current sample of
transiting planets
(e.g.,~\citeauthor{burrows07}~\citeyear{burrows07}).  The reliability
of our temperature estimate is strongly supported by the results of
several independent checks we have carried out (some more constraining
than others), all of which are in good agreement (additional
spectroscopy, line depth ratios, and H$_\alpha$ line profiles). The
old age (similar to the Sun) we infer for \mbox{TrES-2} is also
supported by results from the measurements of the \ion{Ca}{2} activity
level, the lithium abundance, and rotation, which rule out a very
young age for the system. This is consistent with the notion that the
planet's inferred mass escape rate ($\sim 5\times 10^{10}$ g s$^{-1}$,
using the model of ~\citeauthor{lecavelier06}~\citeyear{lecavelier06})
is not very high. The inferred lifetime (tens of Gyr, well outside the
evaporation-forbidden region indicated by Lecavelier des
Etangs~\citeyear{lecavelier06}) is long enough that very efficient
evaporation scenarios (\citeauthor{baraffe04}~\citeyear{baraffe04})
are not likely to hold in the case of the \mbox{TrES-2} system.

We have shown that the best constraint on the radius of the parent
star comes not from the spectroscopically determined surface gravity
(the quantity most often used for this purpose when a trigonometric
parallax is unavailable), but from the photometrically determined
quantity $a/R_\star$, which is closely related to the stellar
density. This quantity is directly measured from the transit light
curve, and depends only very weakly (or in some cases, not at all) on
any assumed stellar properties. In this particular case the gain from
using this new constraint is about a factor of five in terms of the
precision in $R_\star$. Similar arguments for using $a/R_\star$
instead of $\log g$ were made by Pont et al.~\citeyearpar{pont07}
regarding the faint parent star of the transiting planet OGLE-TR-10b,
although they appear not to have actually applied the method in
arriving at the mass and radius of the star reported in their
work. With the significant improvements seen recently in the quality
of the light curves of several of the known transiting planets
(\citeauthor{charbonneau06}~\citeyear{charbonneau06},~\citeyear{charbonneauetal07};
~\citeauthor{bakos06}~\citeyear{bakos06};
~\citeauthor{gillon06}~\citeyear{gillon06};
~\citeauthor{holman06}~\citeyear{holman06},~\citeyear{holman07a},~\citeyear{holman07b};
~\citeauthor{winn07a}~\citeyear{winn07a},~\citeyear{winn07b},~\citeyear{winn07c};
~\citeauthor{knutson07}~\citeyear{knutson07};
~\citeauthor{pont07}~\citeyear{pont07};
~\citeauthor{minniti07}~\citeyear{minniti07};
~\citeauthor{diaz07}~\citeyear{diaz07}), the measured $a/R_\star$
values are likely to be much better now than in the original discovery
papers. Thus, it may pay to revisit the determination of stellar
parameters of many of these systems along the lines of what we have
done here, since this should result in significant improvements in the
absolute mass and radius estimates of the attending exoplanets as
well. Such a study is underway by a subset of the present authors.

Our improved stellar parameters yielding more precise values for the
planet mass and radius confirm that \mbox{TrES-2} is the most massive among
the currently known nearby ($d\lesssim 300$ pc) transiting hot
Jupiters. Although the surface gravity of transiting planets, $\log
g_p$, has not usually attracted much attention, we point out here that
the little-used quadratic relation between the mass and radius of a
transiting planet allows the determination of this property of the
object purely from observations, free from assumptions about the mass
and radius of the parent star. The high accuracy with which $\log g_p$
can be determined in \mbox{TrES-2} and other cases makes it a potentially
useful constraint to theory.

Among the fourteen extrasolar giant planets known to transit their
parent stars, \mbox{TrES-2} seems to belong to an increasing family of
objects (HD~209458b, HAT-P-1b, WASP-1b) whose measured radii
apparently disagree with published theoretical models, in that they
are larger than expected. In the past, this anomaly was explained
invoking a source of internal heat
(e.g.,~\citeauthor{bodenheimer01}~\citeyear{bodenheimer01},~\citeyear{bodenheimer03};
~\citeauthor{guillot02}~\citeyear{guillot02};
~\citeauthor{winn05}~\citeyear{winn05}), either through eccentricity
pumping by more distant companions or through persisting obliquity
tides. However, these scenarios have some difficulties
(\citeauthor{laughlin05}~\citeyear{laughlin05};
~\citeauthor{levrard07}~\citeyear{levrard07}). More recent work
(\citeauthor{burrows07}~\citeyear{burrows07}) indicates that the
discrepancy may be due instead to super-solar metallicities and
opacities that naturally retain internal heat, thus helping to keep a
hot Jupiter's radius larger for longer times. The
smaller-than-expected radii of other transiting planets (e.g.,
HD~149026b) can instead be explained by the presence of dense rocky
cores, and a correlation has been proposed between inner core masses
and host star metallicities
(\citeauthor{guillot06}~\citeyear{guillot06};
~\citeauthor{burrows07}~\citeyear{burrows07}). In this respect, the
very small inferred core mass for \mbox{TrES-2}
(\citeauthor{burrows07}~\citeyear{burrows07}) agrees well with the
lower-than-solar iron abundance measured for the parent star.

\begin{figure}
\plottwo{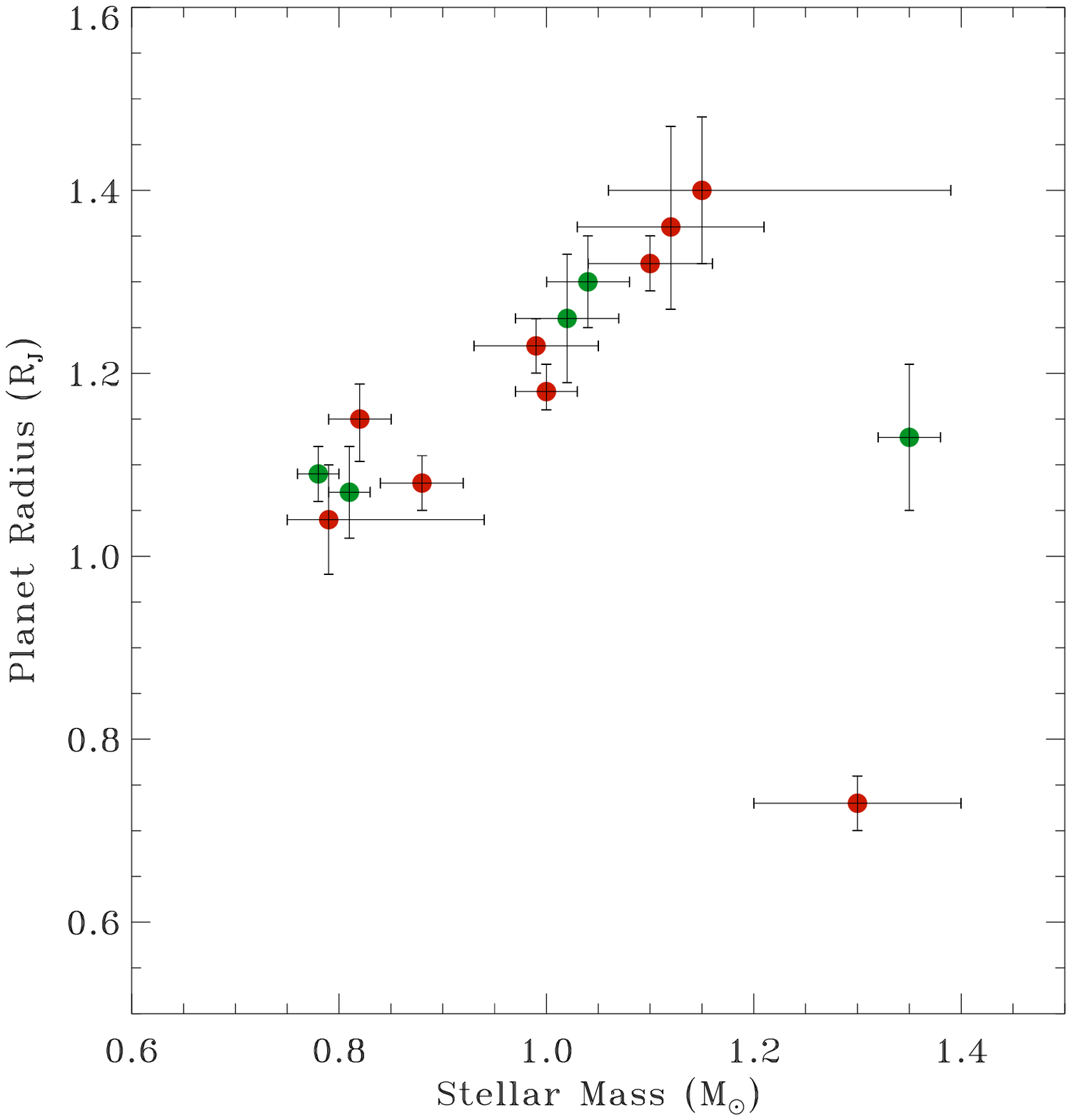}{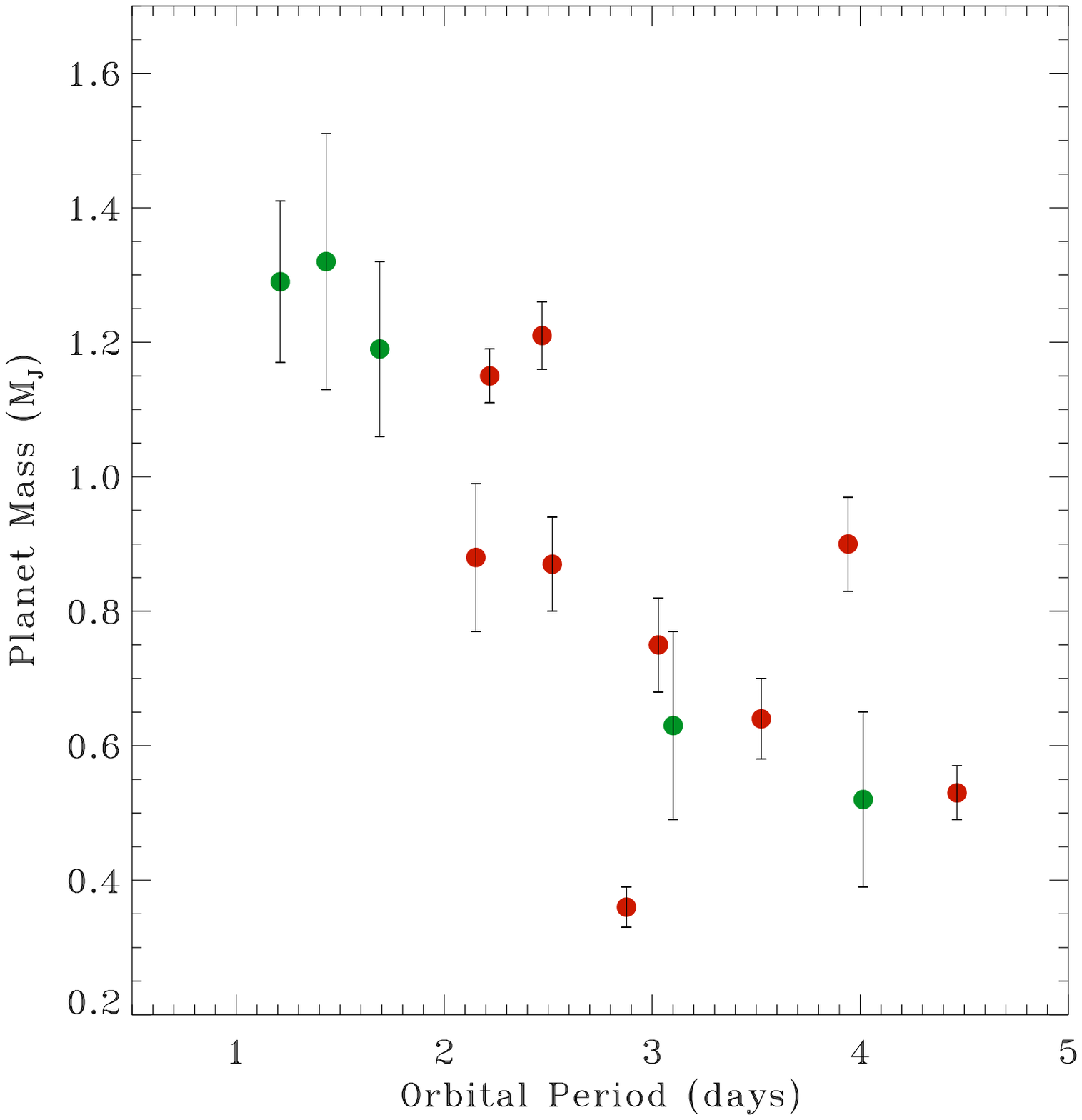}
\caption{Left: Planet radius as a function of host mass for the
fourteen currently known transiting systems. Data are from Burrows et
al.~\citeyearpar{burrows07}, and references therein, except for \mbox{TrES-2} (this
work). Green circles represent the distant OGLE sample, red circles
indicate the sample of transiting objects found orbiting nearby
($d\lesssim 300$ pc) stars. Apart from two outliers, OGLE-TR-132b and
HD~149026b, the correlation between these parameters appears clear in
both the OGLE and the nearby samples of transiting giant
planets. Right: Planet mass as a function of orbital period for the
same sample, with the same color coding. Here the three OGLE planets
with $P < 2$ days drive the correlation (and no planets have been
found yet in this period range by wide-field transit surveys), which
vanishes if the sample of nearby systems only is considered.  It is
still a matter of debate whether the lack of lower-mass planets
($0.5\,M_{\rm Jup}\lesssim M_p\lesssim 1\,M_{\rm Jup}$) with $P < 2$
days could be attributed to uncertainties in the determination of the
stellar (and by inference, planetary) parameters for the faint hosts
(e.g.,~\citeauthor{konacki05}~\citeyear{konacki05};
~\citeauthor{santos06}~\citeyear{santos06};~\citeauthor{pont07}~\citeyear{pont07}),
or whether it could be explained in terms of biases and/or selection
effects (e.g.,~\citeauthor{gaudietal05}~\citeyear{gaudietal05};
~\citeauthor{gaudi05}~\citeyear{gaudi05},~\citeyear{gaudi07};
~\citeauthor{gould06}~\citeyear{gould06}). It is also possible there
might be different upper-mass limits in the two populations, for
example from orbital migration and/or evaporation rate arguments
(\citeauthor{mazeh05}~\citeyear{mazeh05};
~\citeauthor{gaudietal05}~\citeyear{gaudietal05}).
\label{corr}}
\end{figure}

Considering the rapidly growing sample of transiting giant planet
systems as an ensemble, other suggestive, albeit preliminary, trends
between different planet and/or stellar properties have begun to
emerge from the data. Some are easier to explain within the context of
the present theoretical framework, while others still await
explanation.  For example, the apparent trend (with two outliers) seen
in Figure~\ref{corr}, left panel, of increasing planet radius with
increasing mass of the parent star could be due in part to the fact
that close-in planets orbiting more massive stars are more strongly
irradiated (\citeauthor{burrows07}~\citeyear{burrows07}).  On the
other hand, the trend of decreasing planet mass with increasing
orbital period (Figure~\ref{corr}, right panel), first highlighted by
Mazeh et al.~\citeyearpar{mazeh05}, appears to be less well
understood.  We note, however, that in this case the fact that the
host sample is composed of relatively bright, nearby dwarfs targeted
by wide-field transit surveys as well as fainter, more distant OGLE
targets may play some role (see figure caption for details).

The above relations suggest there is a strong interplay between planet
properties and host star characteristics, which is hardly unexpected,
but it is also clear that the parameter space of properties to be
investigated is quite large.  Observations of transiting planet
systems can best inform structural and evolutionary models when they
yield accurate determinations of both planet and stellar properties,
through high-precision photometric as well as spectroscopic
measurements such as those presented here for the \mbox{TrES-2} system.

\acknowledgments

We thank K.\ Biazzo (Catania Astrophysical Observatory) for providing
the LDR calibrations in advance of publication, and A. Burrows and
A. Spagna for helpful discussions. GT acknowledges partial support for
this work from NASA Origins grant NNG04LG89G. DC is supported in part
by NASA Origins grant NNG05GJ29G. AS gratefully acknowledges the
Kepler mission for partial support under NASA Cooperative Agreement
NCC 2-1390. JBL gratefully acknowledges support from NSF grant AST-0307340. 
Some of the data presented herein were obtained at the
W.M. Keck Observatory, which is operated as a scientific partnership
among the California Institute of Technology, the University of
California and the National Aeronautics and Space Administration. The
Observatory was made possible by the generous financial support of the
W.M. Keck Foundation. The authors wish to recognize and acknowledge
the very significant cultural role and reverence that the summit of
Mauna Kea has always had within the indigenous Hawaiian community. We
are most fortunate to have the opportunity to conduct observations
from this mountain. This research has made use of NASA's Astrophysics
Data System Abstract Service and of the SIMBAD database, operated at
CDS, Strasbourg, France.

\clearpage

\begin{deluxetable}{lc}
\tablecaption{Properties of the parent star \mbox{TrES-2}\label{tab:star}}
\tablewidth{0pt} 
\tablehead{\colhead{~~~~~~~~~Parameter~~~~~~~~~} & \colhead{Value}}
\startdata
$T_\mathrm{eff}$ (K)\tablenotemark{a}\dotfill    & 5850~$\pm$~50\phn\phn\\
$\log g$\tablenotemark{a}\dotfill                & 4.4~$\pm$~0.1\\
$\log g$\tablenotemark{b}\dotfill                & $4.426^{+0.021}_{-0.023}$\\
$v \sin i$ (km s$^{-1}$)\tablenotemark{a}\dotfill & 2~$\pm$~1 \\
$\xi_t$ (km s$^{-1}$)\tablenotemark{a}\dotfill   & 1.00~$\pm$~0.05 \\
$[$Fe/H$]$\tablenotemark{a}\dotfill              & $-$0.15~$\pm$~0.10\phs \\
$\langle\log R^\prime_{HK}\rangle$\tablenotemark{a}\dotfill  & $-5.16\pm0.15$ \\
$\log\epsilon{\rm (Li)}$\tablenotemark{a}\dotfill & 2.65 \\
$\rho_\star$ (g cm$^{-3}$)\tablenotemark{c}\dotfill  & 1.375~$\pm$~0.065 \\
$M_\star$ ($M_\sun$)\tablenotemark{b}\dotfill    & $0.980\pm0.062$ \\
$R_\star$ ($R_\sun$)\tablenotemark{b}\dotfill    & $1.000^{+0.036}_{-0.033}$ \\
Age (Gyr)\tablenotemark{b}\dotfill               & $5.1^{+2.7}_{-2.3}$ \\
$M_V$ (mag)\tablenotemark{b}\dotfill             & $4.77 \pm 0.09$ \\
Distance (pc)\tablenotemark{b}\dotfill           & $220 \pm 10$ \\
$U$, $V$, $W$ (km s$^{-1}$)\tablenotemark{b}\dotfill  & [$-$1.81, +0.88, $-$5.51]   \\
\enddata

\tablenotetext{a}{Determined spectroscopically.}

\tablenotetext{b}{Inferred from stellar evolution models using
observational constraints (see text).}

\tablenotetext{c}{Derived observationally.}

\tablecomments{The value adopted for the solar abundance of iron is
$\log(N_\mathrm{Fe}/N_\mathrm{H})_\odot = 7.52$}
\end{deluxetable}

\clearpage

\clearpage

\clearpage

\clearpage

\clearpage

\end{document}